\documentclass[useams,useamsmath,usenatbib]{mn2e}
\usepackage{graphicx}
\usepackage{epsfig}
\usepackage{subfigure}
\usepackage{times}
\usepackage{amsmath}
\usepackage{natbib}

\def\gtrsim{\lower 2pt \hbox{$\, \buildrel {\scriptstyle >}\over
{\scriptstyle \sim}\,$}}
\def\lesssim{\lower 2pt \hbox{$\, \buildrel {\scriptstyle <}\over
{\scriptstyle \sim}\,$}}

\def\km{{\rm\,km}}

\def\yrs{{\rm\,yrs}}

\def\Myr{{\rm\,Myr}}
\def\yr{{\rm\,yr}}

\def\AU{{\rm\,AU}}

\def\Myrs{{\rm\,Myrs}}

\newcommand{\SM}{\rm\, M_{\odot}}
\newcommand{\EM}{\rm\, M_{\oplus}}

\title[Capture of Irregular Satellites via Collisions]
      {On collisional capture rates of irregular satellites around the 
       gas-giant planets and the minimum mass of the solar nebula
      }
\author[Koch \& Hansen]
   {F. Elliott Koch $^{1,2}$\thanks{E-mail: ekoch@unsw.edu.au} and Bradley M.S. Hansen $^2$ \\
$^1$Department of Astrophysics, School of Physics, 
   University of New South Wales,\\ Sydney, NSW, Australia \\
$^2$Department of Physics and Astronomy, University of California, \\
    Los Angeles, CA, USA}

\begin{document}

\date{Accepted ---. Received ---; in original form --}

\pagerange{\pageref{firstpage}--\pageref{lastpage}} \pubyear{2010}
\maketitle
\label{firstpage}
\begin{abstract} 
We investigated the probability that an inelastic collision of planetesimals within the Hill
sphere of the Jovian planets could explain the presence and orbits of observed irregular satellites.
Capture of satellites via this mechanism is highly dependent on not only the
mass of the protoplanetary disk, but also the shape of the planetesimal size distribution.
We performed 2000 simulations for integrated time intervals $\sim 2$ Myr and found
that, given the currently accepted value for the minimum mass solar nebula and planetesimal number density 
based upon the \citet{Nesvorny2003} and \citet{Charnoz2003} size distribution $dN \sim D^{-3.5} dD$, 
 the collision rates for the different Jovian planets range between $\sim 0.6$ and $\gtrsim 170 \, \Myr^{-1}$
for objects with radii, $1 \, \km \le r \le 10 \, \km$. 
Additionally, we found that the probability that these collisions remove 
enough orbital energy to yield a bound orbit was 
$\lesssim 10^{-5}$ and had very little dependence on the relative size of the planetesimals. 
Of these collisions, the collision energy between two objects was
$\gtrsim 10^3$ times the gravitational binding energy for objects with radii 
$\sim 100$ km. We find that, capturing irregular satellites via collisions between
unbound objects can only account for $\sim 0.1\%$ of the observed population, hence 
can this not be the sole method of producing irregular satellites.
\end{abstract}  
  
\section{Introduction}  
\label{sec:Intro}  
The irregular satellites of the Jovian planets have been the subject
of much debate since their discovery in the early 1900's.  Since then
there have been approximately 60 found around Jupiter, about 60 around
Saturn, 20 around Uranus and 10 around Neptune 
\citep{Gladman2001, Holman2004, Sheppard2006}.
\citet{NASA_IrregSats} present a current and detailed compilation of the irregular
satellites (e.g. size, orbital characteristics, etc.) for the gas giant planets as 
well as a list of references containing these
data.

Although a clear definition of an irregular satellite remains to be determined,
we use the same definition as \citet{Nesvorny2003}, moons that are
far enough from their parent planet that the precession of their orbital
plane is primarily controlled by the Sun.  As stated in \citet{Nesvorny2003},
this definition of an irregular satellite excludes Neptune's moon Triton.  
The origin of these irregular satellites has yet to be clearly understood
and can give an insight to the formation of our Solar system.  Traditional
theories of planet formation and evolution fail to produce the observed
large orbital inclination observed for many of these satellites, some of which even follow
retrograde orbits with respect to the orbital plane of the planets.
\citet{Jewitt2005} review the possible modes that possible modes of 
capturing irregular satellites by the giant planets are;
\begin{itemize}
		 \item {\bf Gas Drag:} Similar to the Sun and Solar nebula,
		 the gas giants are believed to be the result of the gravitational
		 collapse of a planetary gas nebula.  Therefore, it would be
		 possible for irregular satellites to have enough energy 
		 dissipated by the planetary nebula via gas drag 
		 to remain gravitationally bound to the planet.  \citet{Pollack1979}
		 model these phenomena as a possible explanation for the capture
		 and fragmentation of Jupiter's prograde and retrograde irregular
		 satellites, but don't provide a clear explanation for the 
		 irregular satellites orbiting Uranus and Neptune.  Furthermore,
		 with out the dissipation of planetary nebula, once captured, the
		 captured object would spiral into the planet $\sim 10$ yrs
		 \citep{Nesvorny2003}.
		 \item {\bf Pull Down:} Pull down is the sudden mass-growth
		 of the parent planet thereby capturing the satellite
		 \citep{Heppenheimer1977}. This mode of capture occurs when
		 an object enters the Hill sphere (region around a planet with
                 radius, $R_H = a (m_p/\SM)^{1/3}$, where
                 the Sun's gravity is negligible compared the planet's) through a Lagrange point
		 and while at this semi-stable orbital point ($\sim 100 \yrs$),
		 the mass of the parent planet increases enough for the object
		 to become gravitationally bound to the planet.
		 \item {\bf Collisions:} Gravitational three body interactions
		 (both colliding and non-colliding) occurring within the 
		 Hill sphere of the planet could lead to the capture of one
		 of the objects as well as fragments resulting from high energy
		 collisions \citep{Colombo1971, Weidenschilling2002}.  
		 \citet{Nesvorny2003} calculate the collision rate between
		 the existing irregular satellites of Jupiter to be $\lesssim 1$
		 every 4.5 Gyr, implying that their formation hardens back to a 
                 time when the Solar system was much denser.
\end{itemize}

The purpose of this paper is to expand upon the last scenario, which
has achieved some popularity in recent years. 
\citet{Nesvorny2004} that showed that the spectral characteristics of 
the irregular satellite groups around Jupiter and the other planets were surprisingly similar.
This has been taken to suggest that they are all drawn from the same underlying progenitor
population, and captured during the early dynamical evolution of the solar system.
The most detailed scenario for such an evolution is the so-called 
Nice model for the solar system origin \citep{Nesvorny2007, Bottke2010}.
The initial conditions in this model are chosen to meet specific benchmarks of a larger scenario
involving the explanation of various dynamical structures in the Solar system, and
do seem to provide a path to capturing irregular satellites around some of the giant planets.
Nevertheless, it is also useful to isolate the conditions necessary for a specific observable.
 Thus, in
this paper we wish to evaluate the requirements of a simpler model, in which we 
determine the conditions that allow
the irregular satellite populations to result from collisional interactions within the
tidal field of a giant planet. 
This will also allow us to estimate the size of nebula that would match
conditions necessary to the in situ formation of Uranus \& Neptune proposed by \citet{Goldreich2004a}.
If irregular satellites could be the remnants of collisions
that were captured by the planet as a result of said collision, collision rates and probabilities of
remnants being captured by the planets could be used to set limits on
the minimum mass Solar nebula (MMSN) as well as sizes and number densities of 
planetesimal seeds that eventually formed the inner planets, asteroid
belt, Edgeworth-Kuiper belt objects and the Oort cloud.

The following section, \S~\ref{sec:EarlySS}, briefly describes the epoch in the
formation of the Solar system when the irregular satellites were most likely to
have been captured. \S~\ref{sec:Relavance} discusses why it is important for us
to understand how these satellites were captured by their host planets. The 
presentation of how the data were generated, the motivation behind it
and justification for it are contained in \S~\ref{sec:NbodyAnal}, which is
followed by a detailed description of how various probabilities were determined
leading up to the calculation of the probability that a mass remains bound to 
a planet, \S~\ref{sec:ProbBound}. The paper concludes with the results of these 
analyses, \S~\ref{sec:IrregResults}, conclusions that can be drawn from these 
results, \S~\ref{sec:Conc}, and suggested future work, \S~\ref{sec:Future}.
\section{The Early Solar System: Seeds for Irregular Satellites}
\label{sec:EarlySS}
\citet{Goldreich2004a} and references therein give a detailed description of the early
stages of planetary formation of our Solar system. The era most important to
this investigation is that of oligarchy since this represents the time just prior to the
formation of the inner terrestrial planets. At this point, the cores and the gas
envelopes of the gas giant planets have already formed.
Oligarchy is the point in the evolution of the Solar system where the formation
of planets transferred from ``runaway'' growth to ``ordered'' growth 
\citep{Kokubo1998}.  During this epoch, the planets in the outer Solar system had
essentially finished forming, while the terrestrial planets were still acreting 
material.  {\em N}-Body simulations presented in \citet{Goldreich2004b} and references therein show
that for this to occur, the Solar nebula must be approximately equal in mass to the minimum
mass in the inner Solar system and about six times the minimum mass of the Solar 
nebula in the outer solar system.  This distribution of material is required to 
ensure the formation of Uranus and Neptune in their current location within the age
of our Solar system (the actual surface densities used were $\sigma = 7 $g cm$^{-2}$
at 1 AU and $\sigma = 1.5$ g cm $^{-2}$ at 25 AU) \citep{Goldreich2004a}.  

The epoch directly following oligarchy and thought to be the longest period in
the evolution of our Solar system is ``clean-up''.
At the beginning of this phase of the evolution of our Solar system Jupiter and Saturn
are likely to have been completely formed, Uranus and Neptune almost completely so.
It is likely that $\gtrsim 1000 \; {\rm M}_{\earth}$ of small bodies were distributed 
amongst annulii with gravitational unstable orbits, to account for the expected 
distribution of $\sim 100 \; {\rm M}_{\earth}$ of small
bodies ejected by Uranus and Neptune and a current estimate of  $ 1-10 \; {\rm M}_{\earth}$ 
for the mass of the Oort cloud \citep{Goldreich2004a}.  

\section{Irregular Satellites: A Window to the Past}
\label{sec:Relavance}
As stated in the first \S~\ref{sec:Intro}, the irregular moons of the giant planets, like Edgeworth-Kuiper belt objects,
are thought to be remnants of the early Solar system.  With this in mind, an accurate
description of their ``capture'' by their host planet and possibly the evolution of 
their bound orbits will provide insight to conditions in early Solar system.

As stated in \S~\ref{sec:Intro}, it is difficult to produce a model where the irregular satellites
form from the same circumplanetary disk from which the regular satellites formed, implying that
the irregular satellites were captured by the planet through some other means.  For these satellites to be captured
through via collisions, the objects collide 
within the Hill sphere of the planet and this collision must dissipate enough energy to result in 
at least one of objects remaining gravitationally bound to the host planet. Therefore, we
need to determine the probability that a planetesimal is located around the planet, then a distribution
of where around the planet this object is most likely to be encountered by a second object.  The probability of two objects
colliding can be calculated from the product of these two distributions and a ratio of cross sectional areas 
(\S~\ref{sec:IrregResults}). Finally, a collision rate can be calculated by approximating the number
of planetesimal present near the planets from accepted values of the minimum mass Solar nebula and 
assuming the density of these objects is comparable to that of the Earth.
%
\section{Integration and Analysis}
\label{sec:dataGen}
Expanding on the principle that these ``small bodies'' only weakly gravitationally interact with each other
we chose to perform multiple simulations consisting of the Jovian planets, the Sun
and only one small body,  
which is equivalent to a {\em N}-body integration containing $\gtrsim 2000$
objects.  Ideally we would have populated the orbital annuli gravitationally stirred by the
planets with an oligarch number density equivalent to $\gtrsim 100 {\rm M}_{\oplus}$, but 
these numbers were computationally prohibitive. To avoid this complication, we assumed
that oligarch-oligarch interactions can be ignored for {\em N}-body simulations run
at the Solar system level due to their small relative mass. This allowed us to
performed a large number of integrations containing only one ``small body'' following which we 
based the remaining analysis on probabilities.
At the time of analysis, this methodology was preferred over a smaller number of simulations
with a large number of bodies because parallel processing code to perform these
operations was still being developed.

\subsection{{\em N}-Body Integration}
\label{sec:IrregSatInt}
{\em N}-body integrations were perform using a Burlisch-Stoer integration routine
embedded in the {\em Mercury6} program written by \citet{Chambers99}.  The simulation
integrates the orbital motions of the Jovian planets and one test mass for up to
$8 \times 10^6$ yrs or until the test mass either collides with a ``big body'' 
(planet or the Sun) or is ejected from
the Solar system ($\gtrsim 100 \AU$ from the Sun).  The accuracy used to define the
time step for the integration was defined to be $\epsilon = 10^{-15}$ to ensure accurate results
calculated on a reasonable time frame.  Using this parameter, integration time steps
are determined by \citet{NumRecipes},
\begin{equation}
   H_k = H \left(\frac{\epsilon}{\epsilon_{k+1,k}}\right)^{1/(2k+1)}
   \label{eq:Accuracy}
\end{equation}
where, $H$ is the user defined time step and $\epsilon_{k+1,k}$ is the percent difference
between the $k$ time steps and $k+1$ time steps to span $H$.
Mercury{\em 6} provides tools to monitor ``close encounters'' between the different objects
given the two objects pass within a user defined distance \citep{Chambers99}.  The distance defined to be the
minimal distance defining a ``close encounter'' was a half of a Hill radius,
\begin{equation}
   0.5 R_{H} = 0.5 \, a_p \left(\frac{M_p}{\SM} \right)^{1/3}
\end{equation}
Originally the Mercury{\em 6}
algorithm returned the orbital parameters of the two objects that pass within the predefined
distance, but we altered the code to return the position and velocity vectors of the
small mass measured with respect to the planet, which were then stored for further analysis.

\subsubsection{Initial Conditions}
\label{sec:ICs}
The orbital parameters used to define the initial orbits of the small bodies were all
generated randomly with the exception of the eccentricity, which is initially chosen to be
equal to zero because this is the parameter most sensitive to gravitational perturbation
and had quickly become randomized during the simulation.  In order to reduce the number
of simulations that led to stable orbits that do not interact with any of the planets,
the initial semi-major axis is generated by randomly choosing an orbital annulus that \citet{Ito2002}
have shown to produce unstable orbits in our Solar system and then equated to a random number
that would lie within one of these rings.  Other than the argument of perigee and the ascending
node, which were chosen to be randomly generated numbers between 0 and $2 \pi$, the angle
of inclination was a randomly generated number between $\pm 0.01 ^{\circ}$.  Orbital inclinations
within this range are well within 
the angular thickness of the Solar nebula after it began to ``flatten'', allow for orbits crossing 
the planets beyond their orbital planes, and are low enough to ensure that the small bodies 
had orbits crossing the planets. Although planetesimals could have initially had large orbital inclinations,
objects with inclinations much larger than those chosen were less likely to enter a planet's 
Hill sphere. Furthermore, this range of inclinations still allows objects to pass through a planet's 
Hill sphere with essentially a uniform distribution as expected.
We found (as expected) that the probability of locating an object in a ``bin'' or region around
the planet, $\mathcal{P}_{bin} = \langle t_{bin} / t_{tof} \rangle$ has essentially no radial
nor angular dependence.

\subsection{Analysis of {\em N}-Body Integration}
\label{sec:NbodyAnal}
Following the numerical integration performed with {\em Mercury6}, a more detailed analysis
of the encounters between the small body and planet(s) was performed.  {\em Mercury6} generated
vast quantities of data consisting of the initial positions and velocities of the small body with respect
to the planet whenever the small body passed within $ 0.75 \, R_{H}$ of the planet.  These
data were then used to calculate the fractional time spent within any given volume element, $dV$,
\begin{displaymath}
   dV = \prod_{i=1}^3 \Delta x_i
\end{displaymath}
around the planet.  Due to computational limitations (i.e. RAM as well as CPU capacity), $dV$
was restricted to be equal to $\sim 10^{-7} R_{H}^3$ corresponding to a Cartesian grid from
$-0.75 \, R_H$ to $0.75 \,  R_H$ in each direction consisting of 100 bins or divisions setting
$\Delta x = 0.15 \, R_H$.

The time, $dt$, spent in each bin for a given pass through the Hill sphere is
\begin{equation}
   dt = \frac{ds}{v}
\end{equation}
where $ds$ is the path of the small body through the volume element and both $d\mathbf{s}$ 
and $\mathbf{v}$ point in the same direction.
$dt_i$ is approximated by assuming that the small body is moving along a straight line through
the bin and would therefore be equal to 
\begin{equation}
   dt_i = {\rm\,min} \left(\frac{\Delta x_i}{v_i}\right)
\end{equation}
where $\Delta x_i$ is the distance to the edge of the volume bin in the $i^{{\rm\,th}}$
direction where, $i = 1, 2, 3$ represent the traditional Cartesian directions, $x$, $y$ and $z$ respectively.

In order to determine the probability of a small body being in the $j^{{\rm\,th}}$ bin, $dt$
is then scaled by the time-of-flight of the small body across the Hill sphere, or the amount of time required for the
small body to pass through our region of interest around the planet ($\sim 0.75 R_H$),
\begin{equation}
   t_{tof} = \sqrt{\frac{a^3}{\mu}} \left(e \sinh F - F \right)
\end{equation}
where $F$ is the hyperbolic eccentric anomaly,
\begin{displaymath}
   \cosh F = \frac{e + \cos \varphi}{1 + e \cos \varphi}
\end{displaymath}
and $e$ is the eccentricity, $a$ is the semi-major axis and $\varphi$ is the angular measure from 
periapse.  The ratio of $dt$ to $t_{tof}$ provides the fractional time spent per encounter in any given volume bin.

\subsubsection{Regularly Spaced Propagation of Trajectory for to Determine Where Objects are 
 Likely to be Found within the Hill Sphere for Unbound Trajectories}
\label{sec:Calc_dt/tof}
The trajectory of the small body through a planet's Hill
sphere was calculated from the angle measured from perigee, $\varphi$, and performing a coordinate transformation
from the orbital coordinates, $x^{\prime}$, $y^{\prime}$, to a coordinate system fixed to the planet where
the $x-y$ plane is at the planets equator and the $x$ axis is directed in the same direction as the
$x$ axis use for the Solar system.  Mathematically, the $x-y$ components of the position and
velocity vectors in the orbital plane are;
\begin{eqnarray}
   x^{\prime} &=& r \cos \varphi  \\
   y^{\prime} &=& r \sin \varphi  \\
   v_x &=& -\frac{na}{\sqrt{1-e^2}} \sin \varphi \\
   v_y &=& \frac{na}{\sqrt{1-e^2}} \left(e + \cos \varphi \right)
\end{eqnarray}
where
\begin{equation}
   r = \frac{a \left(1 + e^2 \right)}{1 + e \cos \varphi}
\end{equation}
and  $a$ is the semi-major axis, $e$ is the eccentricity of the small bodies trajectory past the
planet calculated from the initial position and velocity of the small body as it enters
the sphere of interest, and $\mu = n^2 a^3$.  This vector is then rotated to determine where in the planet's fixed reference
frame the object is located.
\begin{equation}
   \mathbf{X}= \mathbf{\mathcal{R}}(\Omega)_{z^p} \mathbf{\mathcal{R}}(i)_{x^{\prime}} \mathbf{\mathcal{R}}_{z^{sb}} \mathbf{X}^{\prime} 
   \label{eq:rotation}
\end{equation}
where $\omega$ is the argument of perigee, $\Omega$ is the ascending node and $i$ is the angle of inclination and
$\mathbf{\mathcal{R}}$ are the rotation matrices around the $z$ axis in the small body's orbital plane, an intermediate
$x$ axis and the $z$ axis in the planet's orbital plane given by;
\begin{eqnarray}
   \mathbf{\mathcal{R}}(\Omega)_{z^p} &=&
    \left(
    \begin{array}{ccc}
      \cos \Omega  & \sin \Omega & 0 \\
     -\sin \Omega  & \cos \Omega & 0 \\
           0       &      0      & 1 
    \end{array} \right) \nonumber \\
   \mathbf{\mathcal{R}}(i)_{x^{\prime}} &=&
    \left(
    \begin{array}{ccc}
      1  &    0    &   0   \\
      0  & -\sin i & \cos i \\
      0  &  \cos i & \sin i 
    \end{array} \right) \nonumber
\end{eqnarray}
and
\begin{displaymath}
   \mathbf{\mathcal{R}}(\omega)_{z^{sb}} =
    \left(
    \begin{array}{ccc}
      \cos \omega  & \sin \omega & 0 \\
     -\sin \omega  & \cos \omega & 0 \\
           0       &      0      & 1
    \end{array} \right)
\end{displaymath}
All of the positions and velocities calculated in the orbital frame are transformed using equation (\ref{eq:rotation}),
which are then used to determine $dt$ and which bin it corresponds to.  There were two reasons for 
simulating the orbit in this manner.  The most important reason was that moving a small body along its trajectory
in discrete angular steps that correspond to the angular width of the bins most distant from the planet ensures that
the trajectories are sampled completely.  Since most optimized integration routines will adjust the time step to
ensure that there is a tolerable offset in a conserved quantity like the total energy or angular momentum, it is likely
that some statistical bins located farthest from the planet would be ``skipped'' over.  Therefore the
angular step $\Delta \varphi = 1.5 \times 10^{-2}$ radians was used to correspond to the angular width of a
bin at the farthest distance of interest, $\Delta x / R_{max}$, where $R_{max}=3/4 \, R_H$.  To a lesser extent, there is
some reduced run-time because Mercury{\em 6} writes data to the hard disk at regular intervals, which would then need to
be extracted to determine the trajectories.  

\section{Calculation of Probabilities}
\label{sec:ColProb}
We calculated the probability of a physical collision between planetesimals 
occurring at a given region in space around the planet from
three specific ratios; (1) the fraction of time spent in the area of interest around the planet, (2) the
fraction of time spent per bin around the planet and (3) the fraction of cross sectional areas (i.e. a ratio
of the small body cross section to the cross sectional area of the region of interest).

As stated in \S~\ref{sec:Calc_dt/tof}, the fractional time spent in the $i^{th}$ volume bin is equal to $dt_i / t_{tof}$.
The fraction of time spent around the planet is equal to $t_{{\rm\,planet}}/t_{int}$.  
For each integration performed, the time a small body ``spends'' around the planet was calculated by totaling 
the time of flight for each ``close encounter'' for that integration,
\begin{equation}
   t_{{\rm\,planet}} = \sum_j \left(t_{tof}\right)_i
\end{equation}
{\em Mercury6} was configured to return the total integration time, $t_{int}$, for each run
since the program would exit if the small
body collided was ejected of if it collided with a large body (a planet or the Sun).  

The probability that an object is in a given bin, $\mathcal{P}_{bin}$, provided the object is passing through a
planet's Hill sphere is,
\begin{equation}
   \mathcal{P}_{bin} = \left(\frac{dt}{t_{tof}}\right)_{bin} \Big\langle \frac{t_{{\rm\,planet}}}{t_{int}}\Big\rangle.
\label{eq:Pbin}
\end{equation}
Angular dependencies were neglected because $P_{bin} $ varied by less than 1 \% over $4 \pi$ sr.
Although this conveys useful information, it is difficult to visualize. More information could be
gathered from a temporal distribution along a trajectory to determine where in the Hill sphere
these collisions could occur.  Since, $dt_i / t_{tof}$ has little angular dependence,
we assumed $dt_i / t_{tof}$ to be constant over all angles,
hence an average $dt_i / t_{tof}$ over $4 \pi$ sr would be comparable to averaging multiple
temporal distributions to find $\langle dt/t_{tof}\rangle_r$.
Therefore, the probability density of finding a small body in a given annulus at a given distance 
($r+dr$) per cubic AU from the planet, within the planet's 
Hill sphere is,
\begin{equation}
   \mathcal{P}_r = \Big\langle \frac{dt}{t_{tof}} \Big\rangle_r 
                   \Big\langle \frac{t_{{\rm\,planet}}}{t_{int}} \Big\rangle
\end{equation}
The probability that two objects occupy the same bin along a trajectory through a planet's Hill sphere can 
be found by integrating the probability density $\mathcal{P}_r$ along an object's trajectory. If these objects
were large enough to occupy the entire volume of a ``bin'', this would then be the probability of two objects 
colliding. Unfortunately this is not the case and this integration must therefore be re-scaled by the ratio of the 
two cross-sectional areas, 
\begin{eqnarray}
 \frac{\tilde{\sigma}_{sb}}{\sigma_{bin}} &=& \frac{\pi r_{sb}^2}{\Delta x^2} \nonumber \\
   &=& \frac{80}{9} \frac{r^2_{sb}}{R^2_H} \pi \times 1000
 \label{eq:AreaRatio}
\end{eqnarray}
If there were only two objects in a planet's Hill
sphere at any given time, then we would only need the ratio of $\tilde{\sigma}_{sb}$ to $\sigma_{bin}$.
Since it is likely that there multiple small bodies in the Hill sphere, $\tilde{\sigma}_{sb}$ needs to be
scaled by the number of small bodies present, $N \, \mathcal{P}_{{\rm\,planet}}$, where $N$ is the total number of small 
bodies on the annulus that interact with the planet and $\mathcal{P}_{{\rm\,planet}} = \langle t_{{\rm\,planet}} / t_{int} \rangle$, 
leading to a cross sectional ratio of
\begin{equation}
   \frac{\sigma_{sb}}{\sigma_{bin}} = n(r_{sb}; M_{\rm MMSN}) \mathcal{P}_{{\rm\,planet}} 
      \frac{80}{9} \frac{\pi r_{sb}^2}{R_H^2} \times 10^{3}.
\end{equation}
where $n(r_{sb};M_{\rm MMSN})$ are the number of objects with radius $r_{sb}$ and properly scales the area occupied by 
the ``small bodies''. The total area occupied by the planetesimals will be this ratio, integrated over all radii 
(or at least the radii of interest - 1 to 200 km).
Lastly, all of these pieces need to be brought together to yield the probability that any two small
bodies will collide within the Hill sphere of a planet.
\begin{eqnarray}
   \tilde{\mathcal{P}}_{col}(r) &=& \int \mathcal{P}^2_r dV  \frac{\sigma_{sb}}{\sigma_{bin}} \label{eq:Pcol} \\
    &=& n(r; M_{\rm MMSN}) \int \Big\langle \frac{dt}{t_{\rm tof}} \Big\rangle^2_r dV
         \Big\langle \frac{t_{{\rm planet}}}{t_{\rm int}} \Big\rangle^2 \frac{\pi r_{sb}^2}{\Delta x^2}. \nonumber
\end{eqnarray}
Assuming that the rocky material in the Solar nebula is uniformly distributed, the surface density of this material
for the Solar nebula disk between Jupiter and Neptune is
\begin{equation}
  \sigma_{SN} \sim \frac{100 \EM}{\pi \left(30^2 - 5^2\right)\AU^2} 
  \label{eq:SN}
\end{equation}
The number density of small bodies orbiting the Sun was calculated based upon the size distribution of 
planetesimals presented in \citet{Nesvorny2003} and \citet{Charnoz2003}. The number of small bodies as a function of 
the MMSN was assumed to be
\begin{equation}
  n(r;M_{\rm MMSN}) = \eta(M_{\rm MMSN}) r^{-1.8} 
   \label{eq:N}
\end{equation}
where 
\begin{displaymath}
   \eta = \frac{M_{\rm MMSN}}{\rho_{\oplus}} \left[ \int r^{-1.8} V(r) dr \right] ^{-1},
\end{displaymath}
$V(r)$ is the volume of the planetesimal and $\rho_{\oplus}$ is the density of the Earth. 
Lastly equation (\ref{eq:Pcol}) must be integrated
over all ``small body'' radii and the total probability that two planetesimals collide within a planet's
Hill sphere is
\begin{eqnarray}
  \mathcal{P}_{col} &=& \int \tilde{\mathcal{P}}_{col}(r) dr \label{eq:Pcol_tot} \\
     &=& \frac{M_{\rm MMSN}}{\rho_{\oplus}} \left[ \int \tilde{r}^{-1.8} V(\tilde{r}) d\tilde{r} \right]^{-1}
        \int r^{-1.8} \frac{r^2}{\Delta x^2} dr \nonumber \\
     &\times& \int \Big\langle \frac{dt}{t_{\rm tof}} \Big\rangle^2 \Big\langle \frac{t_{\rm planet}}{t_{\rm int}} \Big\rangle^2 dV
        \nonumber
\end{eqnarray}
where the integration over $dV$ is the volume of the Hill sphere and the integrations over both $r$ and $\tilde{r}$ are 
with respect to ``small body'' radii given by \citet{Nesvorny2003} and \citet{ Charnoz2003} size distribution of 
$dN \propto D^{-3.5} dD$.

\begin{figure*}
 \centering
  \includegraphics[angle=270,scale=0.75]{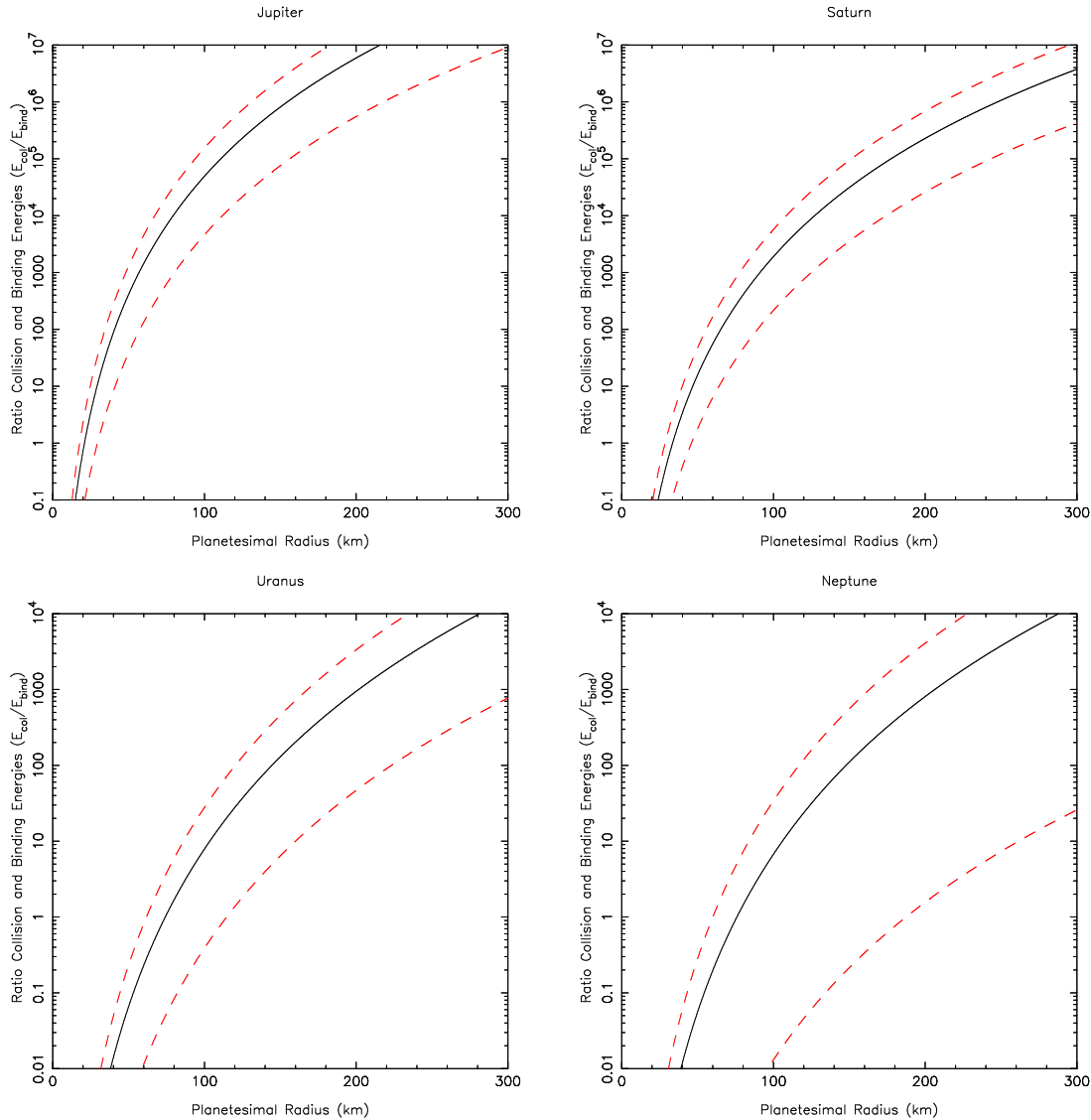}
  \caption{Ratio of collision energy $E_{{\rm col}}$ to gravitational binding energy $E_{{\rm bind}}$
           as a function of the ``small body'' (planetesimal) radius. The (red) dashed lines represent one
           $\sigma$ error for the measured collision energies. The (red) dashed lines correspond to the measured 
           uncertainty of $E_{{\rm col}}$ }
  \label{fig:ColEn}
\end{figure*}
\begin{figure*}
 \centering
  \includegraphics[angle=270,scale=0.75]{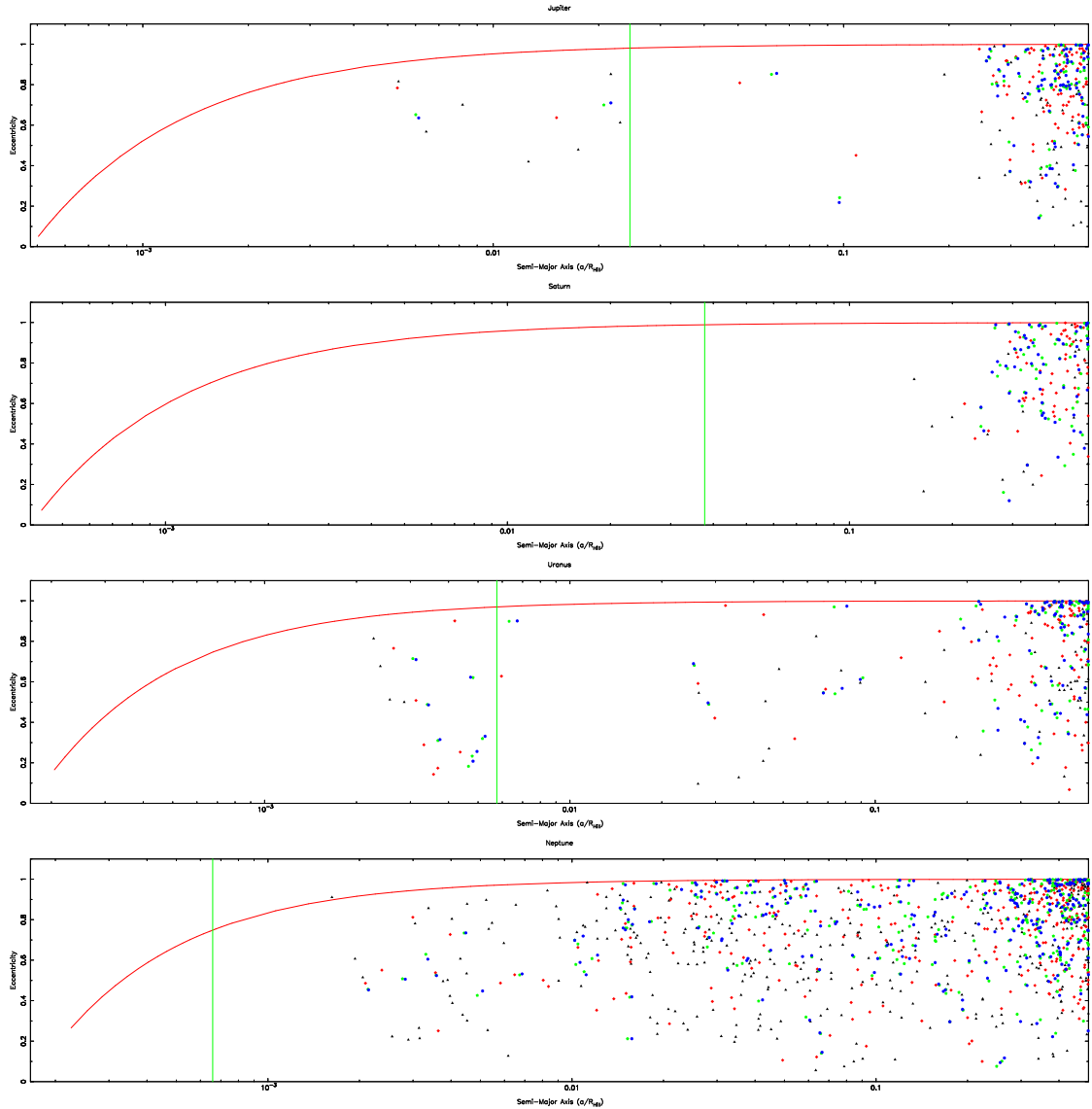}
  \caption{Eccentricity plotted as a function of Semi-Major axis for the conglomerate generated by 
           the collision of two objects within Jupiter's Hill sphere.  The region above the (red) line correspond
           to the bound orbits that pass within Jupiter's radius thereby resulting in a collision
           with the planet and the vertical (green) line the semi-major axis of the ``regular'' moon
           orbiting farthest from the planet.  There are no restrictions placed on the collision energy for the points
           plotted above and the average value for $V_{col}^2$ is $\sim 4 \times 10^{3} \mu_{sb}/r_{col}$,
           the binding energy per unit object of the small body where we assumed $r_{col} \sim 10 \km$. The
           different points correspond to different object ratios between the colliding objects: (black) triangle
           are 1 to 1, (red) diamonds are 10 to 1, (green) pentagons are 100 to 1 and (blue) hexagons are
           1000 to 1. 
           }
   \label{fig:avse_Jup}
\end{figure*}
\subsection{Probability of a Bound Orbit and the Resulting Orbital Parameters}
\label{sec:ProbBound}
The majority of collisions that occurred in a planet's Hill sphere did not result in the object remaining 
gravitationally bound to that planet. The magnitude of the kinetic energy of a typical collision is extremely high
(figure \ref{fig:ColEn}) and is large enough to completely destroy objects with radii $\lesssim 20 \km$. 
Furthermore, even if the two objects were able to coalesce, the majority of collisions did not remove
sufficient orbital energy from the combined object to result in a bound orbit.

However, to simplify matters, we can make some initial  assumptions about the dynamics of a collision needed 
to be made in order to determine which collisions would 
result in a bound orbit or not and revisit these assumptions given the results.  
The first assumption was that the two objects would ``stick'' together, or
\begin{equation}
   \big| \mathbf{v_1} - \mathbf{v_2} \big| < \sqrt{2 G \rho r_{sb}^2}
   \label{eq:NoShatter}
\end{equation}
where the objects have a density, $\rho_{sb}$, equal to the density of the Earth, 
$\rho_{\oplus} \approx 5 {\rm g \,}{\rm cm}^{-3} $. We understand that this is not
the case, but calculating the orbital characteristics of the resulting inelastic collision
is simpler and can serve as a lower limit for the probability of objects being captured.
Although technically not an 
assumption, the trajectories known to pass through a given volume bin ($ d \lesssim 0.01 \AU$, where
$d$ is the distance between the two objects) were ``forced'' to collide
by translating the small bodies to the average position vector with their given velocity.  
This translation results in a error of $\lesssim 1\%$, and the probability of a collision occurring was already
determined in equation \ref{eq:Pcol_tot}. Furthermore the ensemble required to produce even one collision is 
prohibitively large, ``forcing'' trajectories to ``cross paths'' is a reasonable approximation (assumption).
Lastly, provided equation (\ref{eq:NoShatter}) is satisfied, a collision results in a 
bound orbit provided,
\begin{equation}
   V_{cm}^2 < 2 \frac{G M_{{\rm\,planet}}}{r_{col}}
   \label{eq:BoundOrbit}
\end{equation}
where $\mathbf{V_{cm}} = (m_1\mathbf{v_1} + m_2\mathbf{v_2})/(m_1+m_2)$ is the center of mass velocity, 
$\mathbf{r_{col}} = (\mathbf{r_1} + \mathbf{r_2})/2$ is the point where the collision
is ``forced'' to occur and that the agglomerate doesn't smash into the planet (i.e. $a (1-e) > r_{{\rm\,planet}}$).  

Initial conditions for trajectories corresponding with all of the ``close encounters'' for each planet
were stored so that we could use each unique initial condition to integrate trajectories and determine
collision characteristics. Limiting ourselves to 500 unique initial conditions, we assign an initial
condition to one object, then step through the remaining conditions counting the total number of ``assumed''
collisions ($d < \sqrt{2} \Delta x$), $N_c$, and the number that result in a bound orbit, $N_b$ (resulting object
satisfying equation \ref{eq:BoundOrbit}). 
This results in 
\begin{equation}
 \sum_{i=2}^{500} i\left(i-1\right) \sim 4 \times 10^7
 \nonumber
\end{equation}
integrations. These integrations produced $\sim 10^5$ collisions of which $\lesssim 10$ resulted
in a bound orbit (table \ref{tab:ProbBound}).

\begin{figure*}
 \subfigure[Jupiter]{
  \includegraphics[scale=0.4]{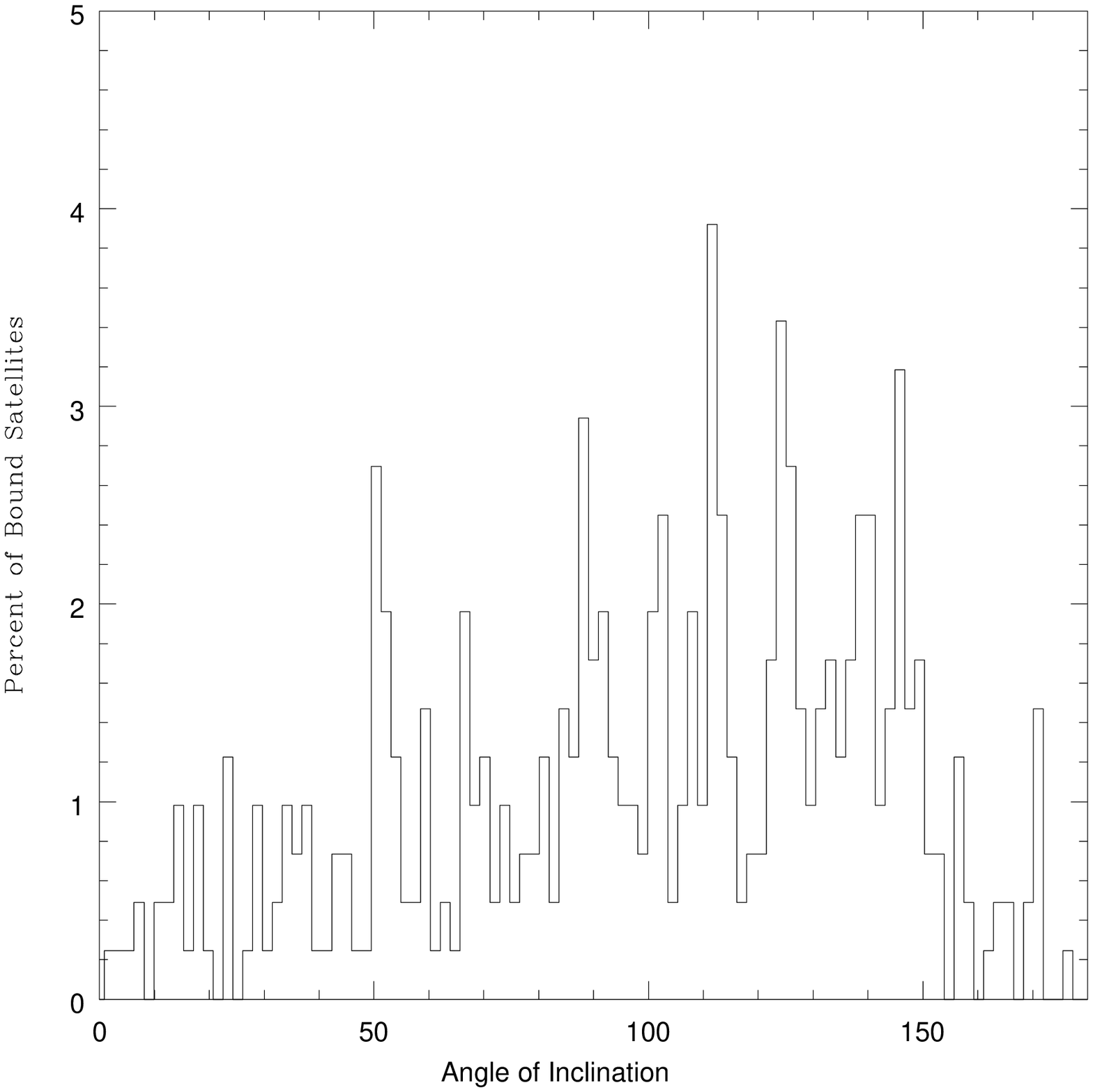}}
 \subfigure[Saturn]{
  \includegraphics[scale=0.4]{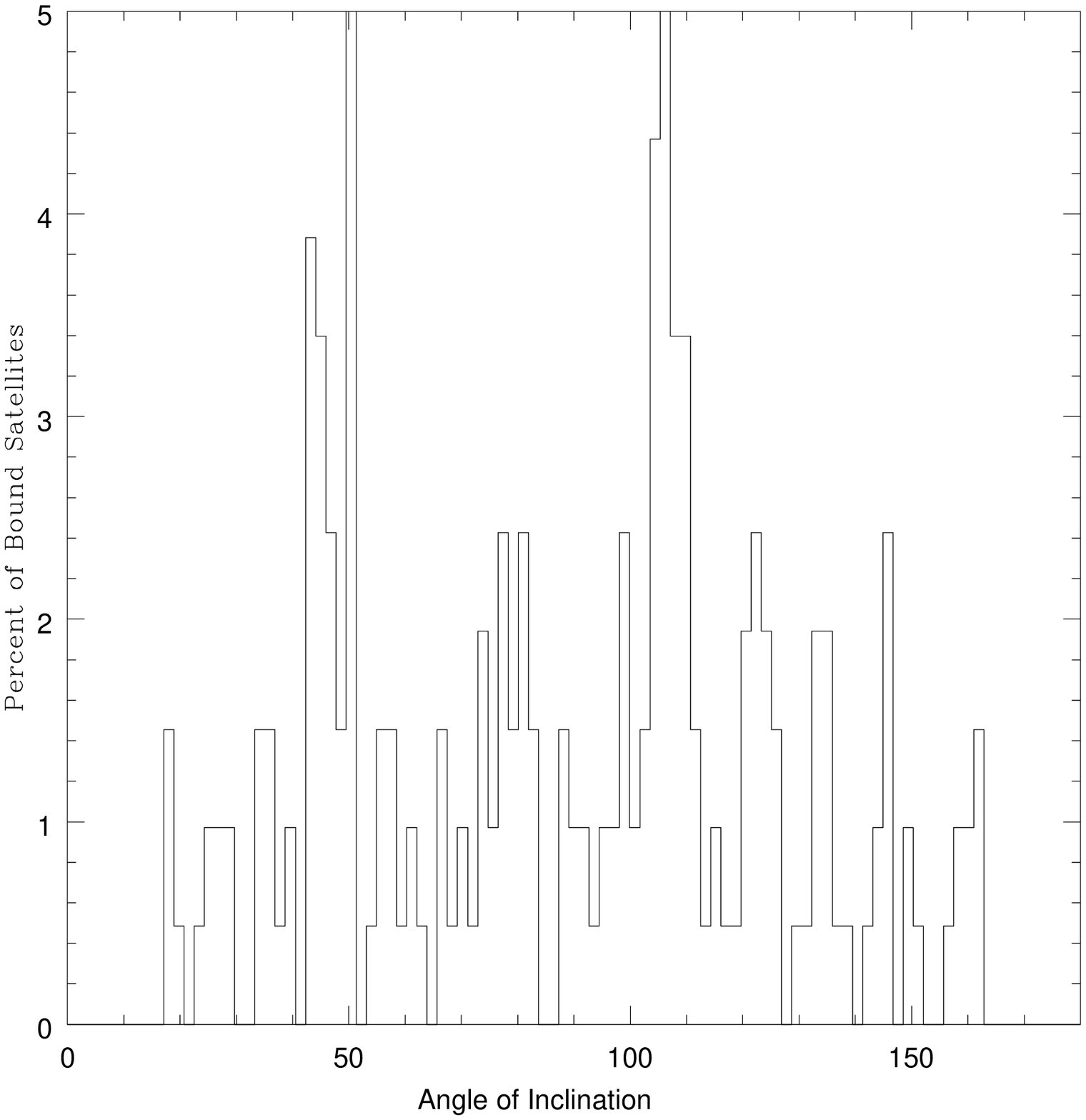}} \\
 \subfigure[Uranus]{
  \includegraphics[scale=0.4]{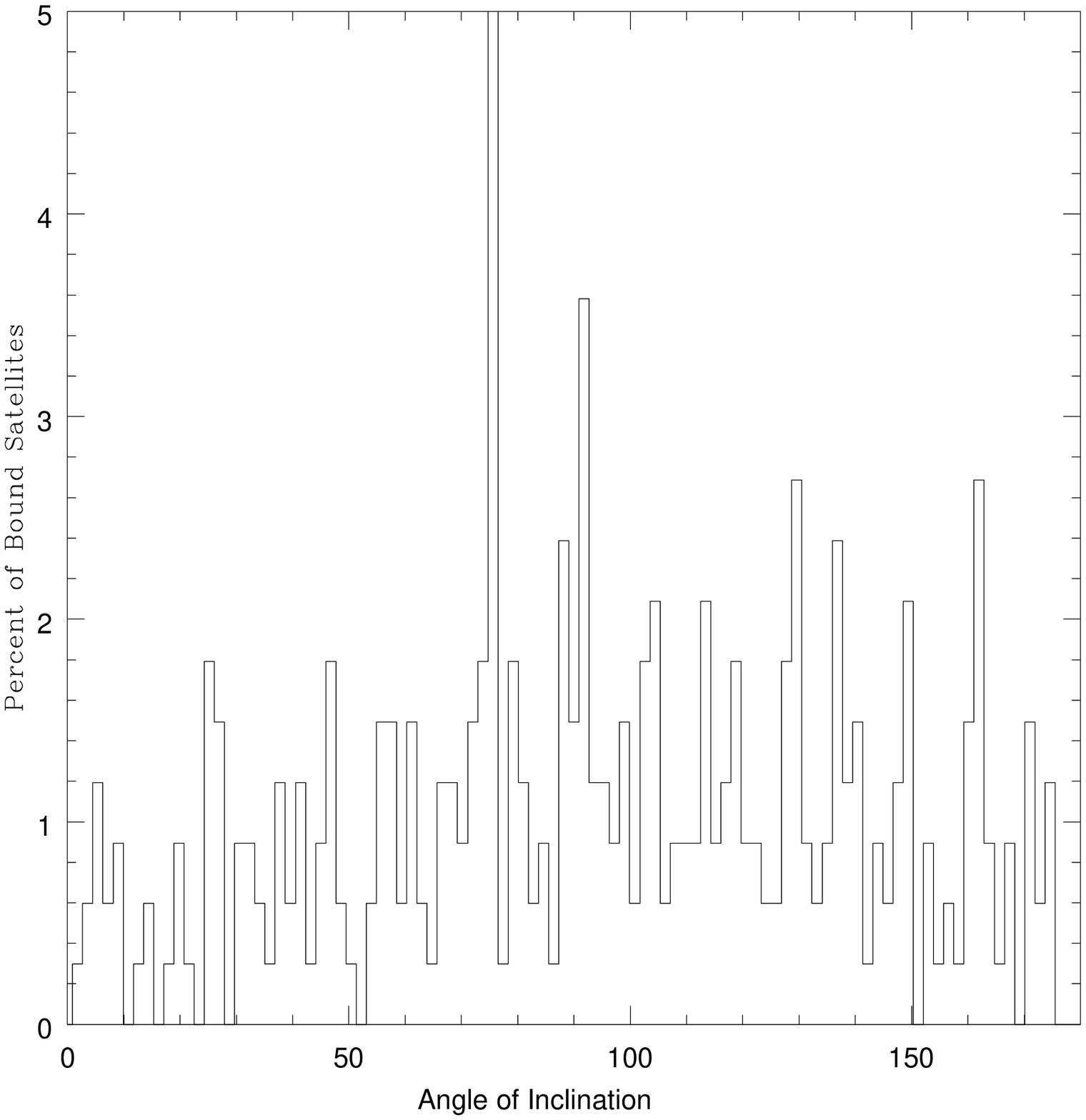}}
 \subfigure[Neptune]{
  \includegraphics[scale=0.4]{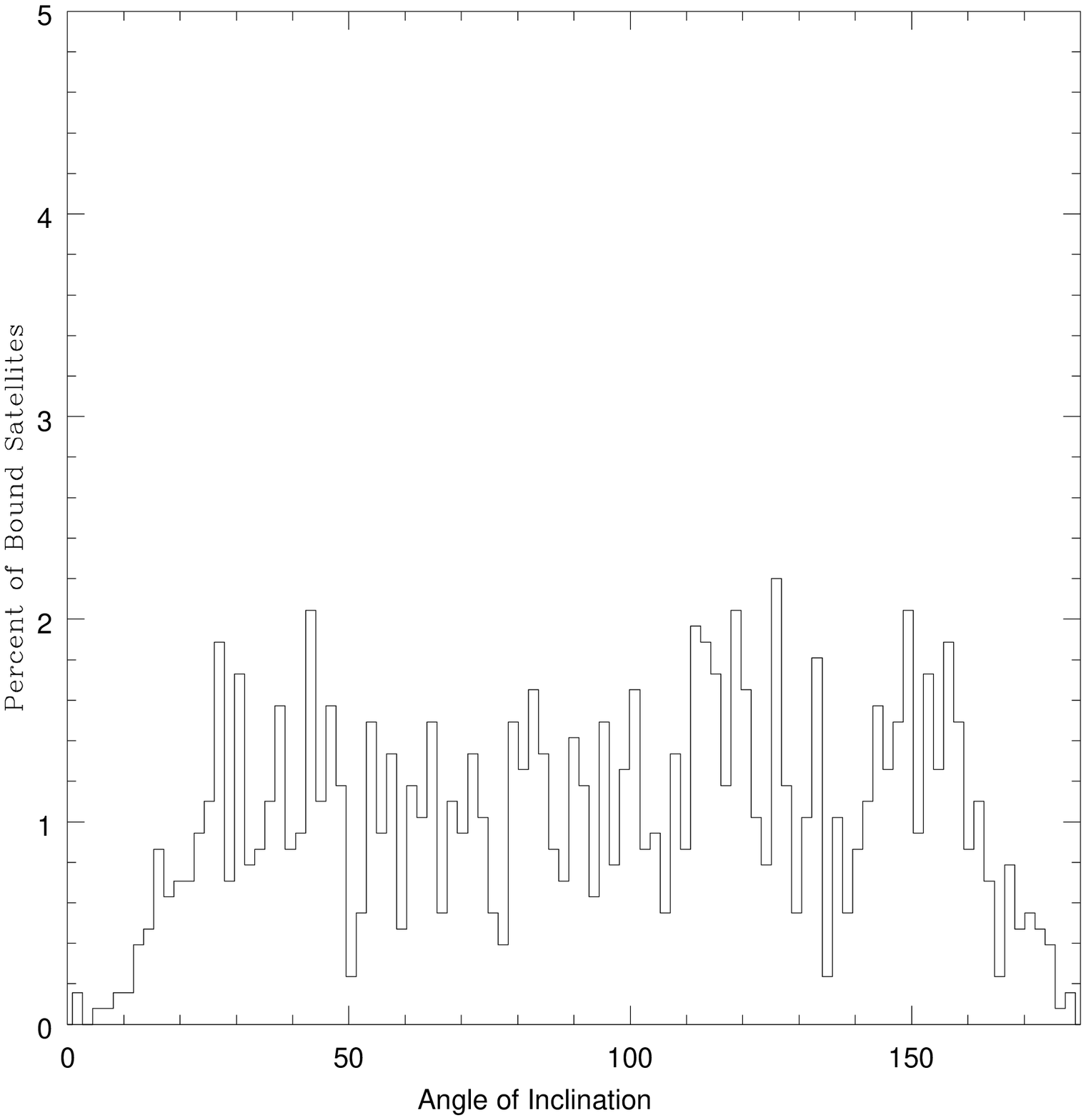}}
 \caption{Distribution of the angle of inclination for the bound orbits represented in figure \ref{fig:avse_Jup}.
   As stated in the assumptions, these inclinations are the result of a collision between to planetesimals within
   a Jovian planets' Hill sphere where the two objects ``stuck'' together.}
 \label{fig:i_dist}
\end{figure*}

Figure (\ref{fig:avse_Jup}) are plots of the eccentricity as a function of semi-major axis for an irregular satellite 
captured by the planets via an inelastic collision between two planetesimal. 
This plot shows that the eccentricities of orbits arising from these collisions are essentially random,
although slightly skewed with a mean of $\sim 0.7$. The
region above the line corresponds to orbits with periapse less than or equal to the 
radius of the planet (in this case Jupiter).  These orbits would result in a collision
with the planet.
Additionally, figure (\ref{fig:i_dist}) shows the distribution of 
inclinations from the resulting bound orbits. These
distributions are essentially random distributions with a couple of the planets having a slightly higher
probability of orbits with marginally retrograde orbits with inclinations $\sim 120 \deg$.

\begin{figure*}
  \centering
  \leavevmode
   \includegraphics[angle=270, scale=0.75]{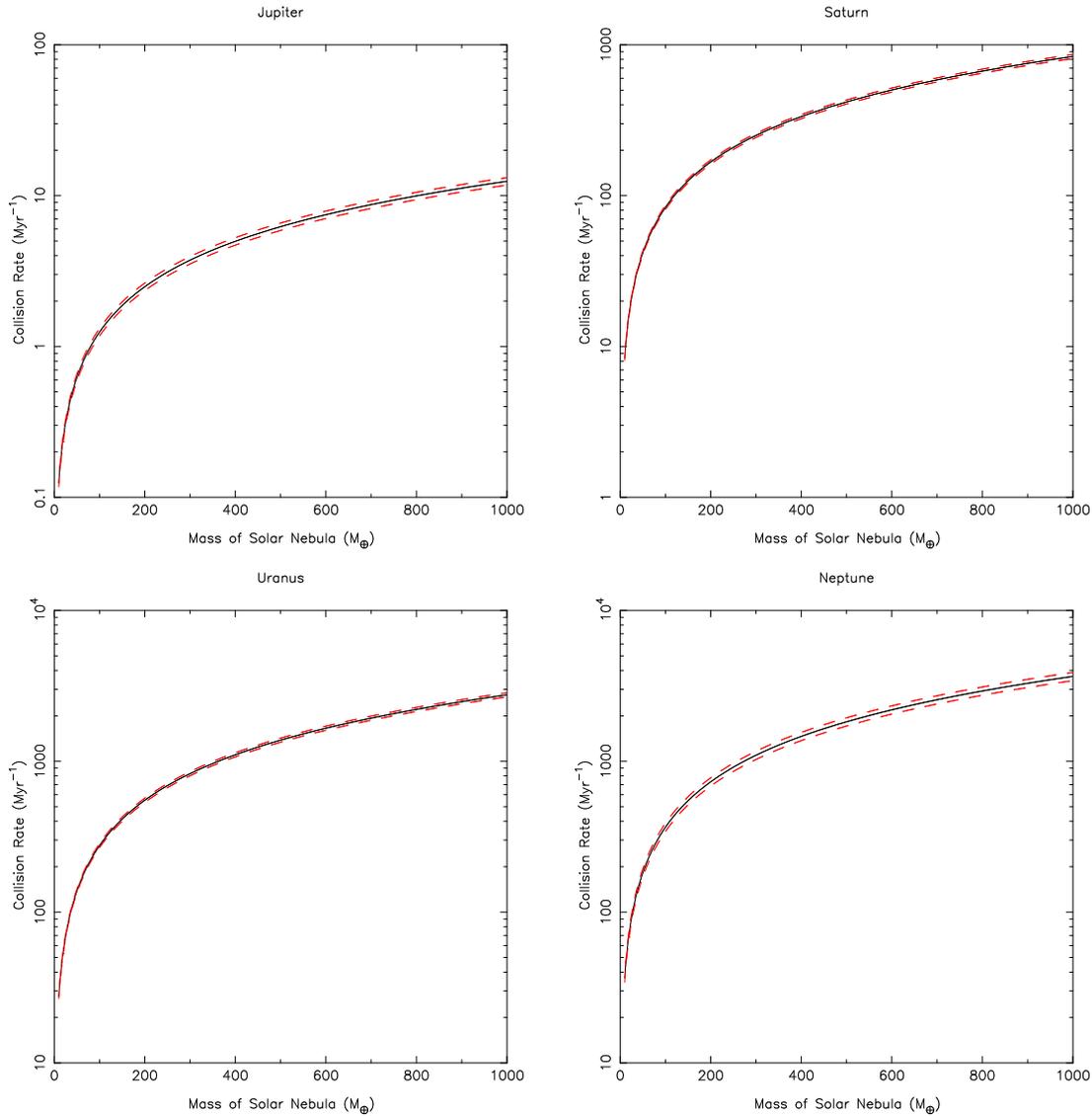}
  \caption{Collision rates for small body collisions occurring within the planet's Hill sphere.
           The number of small bodies from calculated using equation (\ref{eq:N}) is used in
           conjunction with equation (\ref{eq:Pcol_tot}) to calculate collisions per Myr. The (red)
           dashed lines correspond with the propagated measurement uncertainty of
           $\langle dt / t_{tof} \rangle_r$ and $\langle t_{\rm planet} / t_{int} \rangle$}
   \label{fig:ColRate}
\end{figure*}
\section{Results}
\label{sec:IrregResults}
We ran 2000 random simulations resulting in $\gtrsim 2 \times 10^8$ close encounters between a given
planet and a test mass that revealed the mean fractional times, $\langle t_{{\rm\,planet}}/t_{int} \rangle$
 spent around Jupiter, Saturn, Uranus and 
Neptune are $5.48 \times 10^{-5}$, $5.32 \times 10^{-5}$, $1.31 \times 10^{-5}$ and $6.29 \times 10^{-5}$
respectively (described in \S~\ref{sec:ColProb}).  As stated in \S~\ref{sec:IrregSatInt}, numerical 
integrations simulated periods of $\sim 8 \Myrs$ unless the test mass was ejected or collided with either 
a planet or the Sun prior to the 1 Myr period being completed.  Although we attempted to place 
the test masses on known unstable orbits around the Sun determined from \citet{Ito2002}, $\sim 40 \%$ of 
the simulations ran the entire $8 \Myrs$, $\sim 25 \%$ ended $\lesssim 1 \Myr$ and the remaining $35 \%$ 
uniformly distributed between $1-5 \Myrs$.  

\begin{table*}
\caption{Collision Rate of Irregular Satellites (Myr$^{-1}$)}
 \label{tab:ColRate}
\begin{center}
\begin{tabular}{ccccc}
\hline
\smallskip
    Object Radius (km) & Jupiter & Saturn  & Uranus  & Neptune  \\
\hline
\hline
    $ 1 \le r \le 10$ & $0.601 \pm 0.304 $ & $40.30 \pm 1.25 $ 
       & $ 133.1 \pm 4.2 $ & $176.3 \pm 11.0$ \\
    $ 10 \le r \le 100$ & $(1.90 \pm 0.11) \times 10^{-3}$ & $(1.28 \pm 0.04) \times 10^{-1}$
       & $ (4.21 \pm 0.13) \times 10^{-1} $ & $(5.57 \pm 0.35) \times 10^{-1} $ \\
    $ 100 \le r \le 1000$ & $(6.01 \pm 0.34) \times 10^{-7}$ & $(4.03 \pm 0.01) \times 10^{-3}$
       & $ (1.33 \pm 0.04) \times 10^{-3} $ & $ (1.76 \pm 0.11) \times 10^{-3}$ \\
\hline
\end{tabular}
\end{center}
\end{table*}
Using the formulae presented in \S~\ref{sec:ColProb}, we found that the fractional time that a test mass spends around any
given planet was $\sim 10^{-5}$, therefore if an object remained gravitationally bound to the Sun on an unstable
orbit for $1 \Myr$ it would be likely to spend $\sim 50 \yr$ within the Hill sphere of any one of Jovian planets. 
Assuming that the MMSN is $\sim 100 \, \EM$, collision rates for varying object sizes are presented
in table (\ref{tab:ColRate}). For object radii between 1 and 10 km, collision rates are
$0.601 \pm 0.304 \, \Myr^{-1} $
for Jupiter, $40.3 \pm 1.25 \, \Myr^{-1} $
for Saturn $133.1 \pm 4.2 \, \Myr^{-1}$ for Uranus and 
$176.3 \pm 11.0 \, \Myr^{-1}$ (figure \ref{fig:ColRate})
for Neptune. Although these collision rates may appear to be large and imply that there should be a significantly
larger population of irregular satellites, when combined with the probability that a collision between two
planetesimal results in a bound orbit, capture rates for objects with radii between 1 and 10 km are 
(table \ref{tab:CapRate})
$(2.00 \pm 0.22) \times 10^{-5} \, \Myr^{-1} $,
$(6.74 \pm 0.64) \times 10^{-4} \, \Myr^{-1} $,
$(4.19 \pm 0.46) \times 10^{-3} \, \Myr^{-1}$,
$(6.40 \pm 0.52) \times 10^{-3} \, \Myr^{-1}$,
for Jupiter, Saturn, Uranus and Neptune respectively, for a MMSN of $\sim 100 \EM$.
Given energies plotted in figure \ref{fig:ColEn}, it is possible that a collision 
between an object with radius $\sim 10 \km$ collided with one with a radius of $\sim 100 \km$, 
which would have resulted in the larger of the two being broken apart. Another point of
interest is that Jupiter has a ``capture'' rate significantly lower than the other planets.
This is most likely due to the fact that as these objects move closer to the Sun, their
speed increases, thereby requiring more energy to be removed when the two planetesimals
collide in order to capture the satellite.

\begin{table*}
\caption{Capture Rate of Irregular Satellites (Myr$^{-1}$)}
 \label{tab:CapRate}
\begin{center}
\begin{tabular}{ccccc}
\hline
\smallskip
    Object Radius (km) & Jupiter & Saturn  & Uranus  & Neptune  \\
\hline
\hline
    $ 1 \le r \le 10$ & $(2.00 \pm 0.10) \times 10^{-5}$ & $(6.74 \pm 0.64) \times 10^{-4}$
       & $ (6.74 \pm 0.64) \times 10^{-4}$ & $(4.19 \pm 0.46) \times 10^{-3} $ \\
    $ 10 \le r \le 100$ & $(6.32 \pm 0.69) \times 10^{-8}$ & $(2.13 \pm 0.20) \times 10^{-6}$
       & $ (1.33 \pm 0.15) \times 10^{-5} $ & $(2.02 \pm 0.17) \times 10^{-5} $ \\
    $ 100 \le r \le 1000$ & $(2.00 \pm 0.22) \times 10^{-10}$ & $(6.74 \pm 0.64) \times 10^{-9}$
       & $ (4.19 \pm 0.46) \times 10^{-8} $ & $ (6.40 \pm 0.52) \times 10^{-8}$ \\
\hline
\end{tabular}
\end{center}
\end{table*}
%

\section{Conclusions}
\label{sec:Conc}
The Jovian planets are known to have between $\sim 20$ and $\sim 60$ irregular satellites orbiting them
depending on the planet (Jupiter $\gtrsim 60$, Neptune $\sim 10$). Assuming that these
satellites were all captured over a period of 1-2 Myrs, the ``capture rates'' for the planets 
would be $\sim 0.001 \Myr^{-1}$.
In \S~\ref{sec:IrregResults} we present the probability that a collision resulted in a bound orbit  
$P_{\rm bound} \sim 10^{-5}$ per Myr for a total MMSN
containing about $100 \, \EM$ and object radii $1  \km  \le r \le  \km$. 
The collision energy scaled by the binding energy of an object with a 100 km radius (figure \ref{fig:ColEn}) shows
that collisions between the different objects would have to have a mass ratio $\lesssim 0.1$ to ensure that both
weren't completely destroyed and a ratio of $\sim 1$ to be broken apart and not reduced to rubble. 
Using a MMSN based upon asteroid measurements presented
by \citet{Morbidelli2009}, which is $\eta_{MMSN} \sim 10^6 \AU^{-2}$ and size distribution given in
\citep{Nesvorny2003, Charnoz2003}, $dN \propto D^{-3.5} dD$, the rate the Jovian planets could expect to capture
irregular satellites through collisions is $\sim 10^{-3}  \Myr^{-1}$ (figure \ref{fig:ColRate}).
This would be consistent with the \citet{Goldreich2004a} oligarchy models
with the exception that objects primarily consist of $\gtrsim 10-100$ km sized objects, not $\sim 1$ km and 
requiring a slightly larger MMSN in order to populate the Edgeworth-Kuiper belt, asteroid belt and Oort cloud.
Since the probability is highly dependent upon the size distribution of planetesimals as well as the 
mass of the protoplanetary disk, we also investigated size distributions of $dN \propto D^{1.8} dD $ \citep{Bottke2010b},
$dN \propto D^{-2.2} dD$ and $dN \propto D^{-2.8} dD$ \citep{Terai2011} 
all producing similar results for object radii $10  \km \le r \le 100  \km $ (e.g. 
same order of magnitude). However for $1  \km \le r \le 10  \km $, collision rates for
a size distribution $dN \propto D^{-1.8} dD$ is approximately half that for $dN \propto D^{-3.5} dD$ and
for the radii $100 \km \le r \le 1000 \km $ its approximately doubled. 
As the magnitude of the power law increases, naturally the probability of
collisions increases for smaller objects because there are a larger number of them. Furthermore, allowing planets to 
migrate as in the Nice model or simulated by \citet{Lykawka2009}, each planet would sweep through
a larger effective area, thereby marginally increasing the probability of a collision between objects.

The results presented above are subject to some caveats. 
\begin{itemize}
 \item The two
  objects colliding, ``stick'' together and form a single object as a result of the collision, which can not be the case 
  because the collision energies are $\gtrsim 10$ times the gravitational binding energy. This assumption
  can result in an underestimation in the orbital energy removed from the collision and lead to an underestimation
  of the number of bound satellites.
 \item The objects are
  assumed to never achieve a ``quasi-bound'' orbit, which is known to occur where objects can remain
  ``bound'' to the planet for more than 100 years. Although these objects are rare, the amount of time
  spent within the Hill sphere is $\gtrsim 10^4$ times those scattering off of the planet. Furthermore
  since these objects are already on a ``quasi-bound'' orbit around the planet, less energy needs to
  be removed from the collision. Each of the impacts from this assumption, like the first, would lead
  to an underestimate of the number of bound satellites. \citet{Horner2006} have shown that the Centaurs
  can be capture like Trojan satellites and even irregular satellites
  are able to be temporarily captured by Jupiter for $\gtrsim 10^4$ years. Also, in recent years
  a number of comets, Gehrels, 
  Shoemaker-Levy 9 and Oterma \citet{Ohtsuka2008} and references therein. Furthermore, \citet{Horner2010}
  have shown such captures to be quite frequent. Such events could significantly boost our capture rates.
 \item All of the planets formed at there present location. It has already been accepted that, at least,
  Neptune has migrated outward since its formation \citep{Malhotra1993}. As planet migrated away from the Sun, there Hill sphere
  increases, which would mean that for planets moving outward (Neptune and Uranus), their Hill spheres would
  have been over estimated, hence the probability of capture as well. We believe that this over estimation 
  is compensated by the fact that as the planet(s) migrated the annulus swept out by their orbit(s) 
  grows giving them ``access'' to a larger number of objects. However, for planets that migrated inward, their 
  Hill spheres were underestimated, this coupled with the ``growth'' of their orbital annulii would lead
  to an underestimation of the number of bound satellites.
 \item The density of objects colliding is $\sim \rho_{\oplus}$, where in actuality the density is less 
  than the density of the Earth because the Earth and Mercury are the objects with the greatest known 
  density. This will lead to an underestimate in the number of objects (planetesimals), hence produce
  an artificially low collision rate.
\end{itemize}
We found that collisions between objects with a mass ratio as large as 1000 to 1
only marginally reduce the capture rate (probability of bound orbits listed 
in table \ref{tab:ProbBound}). It is possible
for a large mass to collide with a smaller mass and, though the smaller of the two would be destroyed
because the collision is large enough to reduce the small mass to rubble, the larger mass would fare much better and
would most likely simply be broken apart, resulting in some of the debris from the collision remaining bound to the planet. 
Furthermore, since $\sim 1000$ 1 km radius objects are contained in a mass with a 100 km radius,
it is possible that only a small fraction of the remnants from the collision could remain bound to the planet
were the remainder did not have enough energy dissipated by the collision. A detailed analysis of the collisions
of these objects and simulations of the distribution would need to be done to determine what fraction
of these remnants would remain bound to the planets.

We found that the average semi-major axes for bound orbits are essentially randomly distributed between 0.3 
and 2 Hill radii, which is consistent with observation with the exception that many of the objects 
captured in the simulation had highly eccentric orbits and many with a semi-major axis $\gtrsim 5$
Hill radii, which is not what has been observed. This is most likely because these objects
were probably stripped from the planet by the Sun's gravity or interacted with the regular
satellites to produce more circular orbits, closer to the planet. The act of satellites being stripped from
the planets' orbit is yet another factor not accounted for in the analysis above. Since nearly half the 
capture objects have orbits that are on the outer cusp of the Hill sphere, either their orbits would have
to decay, or they would be stripped from the planet. 

%
Although it is unlikely that all of the irregular satellites were captured via collisions with the host planet's
Hill sphere, we have shown that there is a finite probability that irregular satellites can be capture through
collisions, which would serve as a lower limit based on the assumptions that have been made. 
Additionally, the large collision energies that we measured would be consistent with the observations
of \citet{Nesvorny2004} who state that some of Jupiter's irregular satellites can be clustered into specific
groups with similar orbital elements and spectral characteristics. 

\section{Further Work}
\label{sec:Future}
There are several issues that need to be addressed in future work, which can be subdivided into
two different sections.  One section incorporates the assumptions made in the simulations
performed and the other the short comings of the current model.

The conditions modeled in this research most closely mimic those presented by \citet{Goldreich2004b}
where the planets are formed at their current orbital locations.  It differs in the
sense that the energy of planetesimal orbits is not dissipated, ``cooled'', and steadily
increases through interactions with the planets, ``heating''.  \citet{Goldreich2004b} state
that interactions between planetesimal will effectively remove much of this energy by
dispersing it amounts the entire planetesimal population referred to as ``cooling''.
Planetesimal cooling will effectively reduce the energy of these objects and the resulting
collision energy, therefore should increase the probability of a bound state.  

In addition to accounting for orbital ``cooling'', dependence on the migration of planet orbits presented
in \citet{Tsiganis2005} and \citet{Lykawka2009} must be investigated.  \citet{Nesvorny2007} present a mechanism of capture
based upon a three body interaction between to gas giants and a planetesimal passing between
them during a close encounter.  This theory can recreate the observed orbits for irregular satellites
around Saturn, Uranus and Neptune, but fails for Jupiter since there are no known cases of Jupiter encountering
the other planets.  \citet{Nesvorny2007} go on to state that Jupiter would have to capture its satellites 
through some other means.  It would be beneficial to determine if this migration increases collision
rates by increasing the effective area of the feeding zones for the planets, hence the number 
objects passing through the planets' Hill spheres as well as produce bound orbits similar to those
observed by \citet{Nesvorny2007} and comparable to observed orbits.

Lastly, due to the high magnitude of the collision energy and low probability of a collision between two objects resulting
in a bound orbit, it is important study the dynamics of the collisions themselves. Since it is most likely that the objects
are broken apart during the collision, the kinematics of these inelastic collisions and the kinetic 
energy of the remnants should be carefully studied in asteroid families to determine the probability that enough
energy has been removed from the remnants of the collision to remain bound to the planet. It is 
likely that this would result in a much higher probability of bound orbit. 

\section{Acknowledgments}
FEK would like to thank both D. Nesvorn{\'y} and J. Horner for their constructive criticism and 
useful comments regarding the assumptions made, analysis and general methodology.

\bibliographystyle{apj}
\bibliography{thesis}
\label{lastpage}
\end{document}